\documentclass[vecphys]{svmult}

\usepackage{graphicx}        
\usepackage[bottom]{footmisc}

\begin{document}

\title*{A Large-Scale Survey of Neutron-Capture Element Abundances in Planetary Nebulae}
\titlerunning{\emph{n}-capture Element Abundances in PNe}
\author{N.\ C.\ Sterling\inst{1, 2}, Harriet L.\ Dinerstein\inst{3}\and
T.\ R.\ Kallman\inst{1}}
\institute{NASA Goddard Space Flight Center, Greenbelt, MD, 
\texttt{sterling@milkyway.gsfc.nasa.gov, tim@milkyway.gsfc.nasa.gov}\and
NASA Postdoctoral Fellow at the Goddard Space Flight Center.  The NASA Postdoctoral Program is administered by Oak Ridge Associated Universities through a contract with NASA.
\and Department of Astronomy, University of Texas at Austin, \texttt{harriet@astro.as.utexas.edu}}
%
%
\maketitle

\begin{abstract}

We present results from the first large-scale survey of neutron(\emph{n})-capture element abundances in planetary nebulae (PNe).  This survey was motivated by the fact that a PN may be enriched in \emph{n}-capture elements if its progenitor star experienced \emph{s}-process nucleosynthesis during the thermally-pulsing asymptotic giant branch (AGB) phase.  We have measured emission from Se and Kr in over 100 PNe, and use literature data to expand our sample to 120 objects.  [Kr~III]~2.199 and/or [Se~IV]~2.287~$\mu$m were detected in 81 PNe, for a detection rate of nearly 70\%.  We derive Se and Kr abundances or upper limits using ionization correction factors derived from photoionization models.  A significant range is found in the Se and Kr abundances, from near solar (no enrichment), to enriched by a factor of ten.  Overall, 41 of the 94 PNe with derived Se and/or Kr abundances or meaningful upper limits exhibit \emph{s}-process enrichments.  Our survey has increased the number of PNe with known \emph{n}-capture element abundances by an order of magnitude, enabling us to explore correlations between \emph{s}-process enrichments and other nebular and central star properties.  In particular, the Se and Kr enrichments display a positive correlation with nebular C/O ratios, as theoretically expected.  Peimbert Type~I PNe and bipolar PNe, whose progenitors are believed to be intermediate-mass stars ($>3$--4~M$_{\odot}$), exhibit little or no \emph{s}-process enrichment.  Interestingly, PNe with H-deficient [WC] central stars do not exhibit systematically larger \emph{s}-process enrichments than other PNe, despite the fact that their central stars are enriched in C and probably \emph{n}-capture elements.  Finally, the few PNe in our sample with known or probable binary central star systems exhibit little \emph{s}-process enrichment, which may be explained if binary interactions truncated their AGB phases.  We also briefly discuss a new observational program to detect optical emission lines of \emph{n}-capture elements, and new atomic data calculations that will greatly improve the accuracy of \emph{n}-capture element abundance determinations in PNe.

\end{abstract}

\section{Introduction}
\label{sec1}
Planetary nebulae (PNe) are the descendants of low- and intermediate-mass stars ($M=1$--8~M$_{\odot}$), the source of approximately half of the neutron(\emph{n})-capture element (atomic number $Z>30$) nuclei in the Universe.  These elements are created by slow \emph{n}-capture nucleosynthesis (the ``\emph{s}-process'') during the thermally-pulsing asymptotic giant branch (AGB) phase, and are conveyed to the stellar envelope during third dredge-up (TDU).  Free neutrons are released during the interpulse phase by $^{13}$C($\alpha,n$)$^{16}$O --- or $^{22}$Ne($\alpha,n$)$^{25}$Mg in more massive AGB stars --- and are captured by iron-peak ``seed'' nuclei, which undergo subsequent \emph{n}-captures and $\beta$-decays to transform into heavier elements \cite{bgw99, her05}.  The enriched material is discharged into the ambient interstellar medium (ISM) via stellar winds and, ultimately, PN ejection.  At solar metallicity, TDU occurs in stars with initial masses $M>1.5$M$_{\odot}$; in less massive stars, stellar winds reduce the mass of the stellar envelope below the critical mass for TDU (0.3--0.5~M$_{\odot}$; \cite{str06}) before dredge-up can take place.  Hence, stars less massive than 1.5~M$_{\odot}$ (and their PNe) are not expected to exhibit \emph{s}-process enrichments.

Studying \emph{s}-process enrichments in PNe provides unique information that cannot be obtained from spectroscopy of AGB stars.  For example, the lightest \emph{n}-capture elements ($Z=30$--36) and noble gases generally cannot be detected in AGB stars, since strong lines of their neutrals and first ions reside in the UV where cool giant stars emit little flux.  In addition, intermediate-mass stars ($M>3.5$~M$_{\odot}$, hereafter IMS) are difficult to study during the AGB, due to heavy extinction from their dusty, optically thick circumstellar envelopes.  However, these objects are readily observable as Type~I PNe \cite{peim78}.

In this contribution, we present results from the first large-scale survey of \emph{n}-capture elements in Galactic PNe.  We have determined the elemental abundances (or upper limits) of the \emph{n}-capture elements Se ($Z=34$) and Kr ($Z=36$) in 120 PNe, and examine correlations between \emph{s}-process enrichments and other nebular and stellar properties for the first time.  Se and Kr are valuable tracers of \emph{s}-process enrichments in PNe, since they are not expected to be depleted into dust; Kr is a noble gas, and Se has not been found to be depleted in the diffuse ISM \cite{card93}.  A more detailed discussion of the results from our survey can be found in \cite{sdk07} and \cite{sd07}.

\section{Observations and Abundance Determinations}
\label{sec2}

We have observed 103 Galactic PNe in the $K$~band (2.14--2.30~$\mu$m) with the CoolSpec spectrometer \cite{les00} on the 2.7-m Harlan J.\ Smith telescope at McDonald Observatory.  Including additional $K$~band spectra from the literature, our sample is comprised of 120 objects.  We detected [Kr~III]~2.199 and/or [Se~IV]~2.287~$\mu$m in 81 of these objects, a remarkable detection rate considering the low cosmic abundances of Se and Kr ($\sim 2\times 10^{-9}$ relative to H in the Solar System; \cite{asp05}).  These lines, first identified by \cite{din01}, are resolved from other nebular features at our survey resolution of $R=500$, with the exception of H$_2$~$v=3$--2 lines in H$_2$-emitting PNe ($\sim$30\% of our targets).  We removed the contaminating flux from these H$_2$ lines with the aid of high-resolution ($R=4400$) observations and the measured fluxes of other observed H$_2$ lines \cite{sd07}.

Kr$^{++}$ and Se$^{3+}$ ionic abundances were derived for each object, using a 5-level and 2-level model atom, respectively.  To determine elemental Se and Kr abundances, we utilized the ionization correction factors (ICFs) of \cite{sdk07}, which were derived from Cloudy \cite{fer98} and XSTAR \cite{kb01} photoionization models.  The derived Se and Kr abundances are accurate to within a factor of 2--3 for most objects in our sample, taking into account uncertainties in the line fluxes, $T_{\rm e}$, $n_{\rm e}$, and the fractional ionic abundances O$^{++}$/O, Ar$^{++}$/Ar, and S$^{++}$/S (taken from the literature) that are incorporated into the ICFs.

The choice of a reference element, for a metallicity-independent determination of \emph{s}-process enrichments, is particularly important.  For most objects in our sample, we used O as a reference element since its abundance is generally more reliably determined than those of other elements.  However, in Type~I PNe, we found that [Ar/O] is larger by a factor of two relative to non-Type~I PNe.  Since Ar is not processed by PN progenitor stars, this implies that O was destroyed in these objects.  O destruction can occur during hot bottom burning, if the temperature at the base of the convective envelope is high enough for the ON-cycle to be activated \cite{kar06}.  Therefore, we utilize Ar as a reference element for Type~I objects, and use the notation [Se/(O,~Ar)] and [Kr/(O,~Ar)] to underscore our choice of different reference elements for these two subclasses of PNe.

Overall, we find a wide range of Se and Kr enrichment factors.  The average [Kr/(O,~Ar)]~=~0.98, with values for individual PNe ranging from $-0.05$ to 1.89, while the mean [Se/(O,~Ar)]~=~0.31, with extrema at $-0.56$ and 0.90~dex.  In 18 PNe exhibiting both [Kr~III] and [Se~IV] emission, we find that [Kr/Se]~=~0.5$\pm$0.2, in good agreement with theoretical predictions \cite{bus01}.  In 41 of the 94 PNe with derived abundances or meaningful upper limits, Se and/or Kr are enriched by more than a factor of two (the dispersion of light \emph{n}-capture element abundances in unevolved solar-metallicity stars; \cite{trav04}).  We interpret the Se and Kr enrichments as evidence for \emph{in situ} \emph{s}-process nucleosynthesis and TDU in PN progenitor stars.

\section{Correlations}
\label{sec3}

Our survey has increased the number of PNe with determined \emph{n}-capture element abundances by nearly a factor of ten.  This enables us to search for correlations between \emph{s}-process enrichments and other nebular and stellar properties for the first time.  In Table~1, we show the mean values of [Se/(O,~Ar)] and [Kr/(O,~Ar)] for different subclasses of PNe.

\begin{table}
\centering
\caption{Mean Se and Kr Abundances in Different Subclasses of PNe$^{\rm a}$}
\label{tab1}       
%
%
\begin{tabular}{lcccccc}
\hline\noalign{\smallskip}
  & Mean &   & Number of & Mean &  & Number of  \\
Property & [Se/(O,~Ar)] & $< \sigma >^{\rm b}$ & Se Detections & [Kr/(O,~Ar)] & $< \sigma >^{\rm b}$ & Kr Detections \\
\noalign{\smallskip}\hline\noalign{\smallskip}
\multicolumn{7}{c}{\textit{Progenitor Mass}} \\
\cline{1-7}
Type I & $-0.03$ & 0.27 & 12 & 0.09 & 0.14 & 3 \\
Non-Type I & 0.36 & 0.26 & 55 & 1.02 & 0.27 & 30 \\
\cline{1-7}
\multicolumn{7}{c}{\textit{Morphology}} \\
\cline{1-7}
Bipolar & 0.27 & 0.38 & 14 & 0.68 & 0.25 & 8 \\
Elliptical & 0.28 & 0.22 & 28 & 1.09 & 0.38 & 15 \\
\cline{1-7}
\multicolumn{7}{c}{\textit{Central Star Type}} \\
\cline{1-7}
$[$WC$]$ & 0.39 & 0.28 & 16 & 0.90 & 0.34 & 10 \\
Non-[WC] & 0.29 & 0.28 & 39 & 1.05 & 0.32 & 20 \\
Binary & $-0.07$ & 0.12 & 5 & 0.82 & 0.08 & 2 \\
\cline{1-7}
Full Sample & 0.31 & 0.27 & 67 & 0.98 & 0.31 & 33 \\
\noalign{\smallskip}\hline
\end{tabular}
\begin{itemize}
\item[(a)] Only PNe exhibiting Se and/or Kr emission and with determined O and Ar abundances are considered.
\item[(b)] The $< \sigma >$ are mean absolute deviations in the Se and Kr abundances.
\end{itemize}
\end{table}

Theoretically, it is expected that \emph{s}-process enrichments will be correlated with the C/O ratio, since $^{12}$C is transported to the stellar envelope along with \emph{n}-capture elements during TDU \cite{bgw99}.  Strong empirical evidence has been found for such a correlation in AGB \cite{sm90} and post-AGB stars \cite{vw03}.  Indeed, [Se/(O,~Ar)] and [Kr/(O,~Ar)] are positively correlated with the gaseous C/O ratio (Figure~1), with correlation coefficients $r=0.45$ and 0.64, respectively.  Interestingly, Se and Kr enrichments do not increase with C/O as rapidly as Sr, Y, and Zr (observed in AGB stars).  This is likely an effect of the smaller \emph{s}-process yields of Se and Kr relative to these three elements \cite{bus01}.

\begin{figure}[t]
\centering
\includegraphics[height=5.5cm]{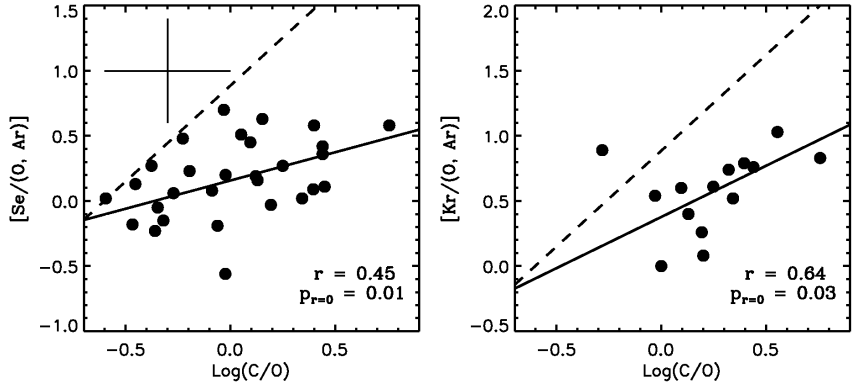}
\caption{[Se/(O,~Ar)] (left) and [Kr/(O,~Ar)] (right) plotted against the gaseous C/O ratio of PNe in our sample.  The best linear fit to each correlation is shown as a solid line (with correlation coefficient $r$ and significance $p_{r=0}$ indicated).  The dashed lines are fits to [$<$Sr, Y, Zr$>$/Fe] as a function of C/O in AGB and post-AGB stars \cite{sm85, sm90, sm93, vw00}.}
\label{fig1}
\end{figure}

We also find that Type~I PNe, which are descendants of IMS, exhibit little \emph{s}-process enrichment compared to non-Type~I PNe (Table~1, Figure~2).  Bipolar PNe also show smaller \emph{s}-process enrichments than elliptical PNe, although the discrepancy is not as pronounced in this case.  This has been verified with Kolmogorov-Smirnov (KS) tests, which show that the probabilities that the Se and Kr enrichments of Type~I and non-Type~I PNe are drawn from the same cumulative distribution functions are $p_{\rm ks}=0.02$ and $p_{\rm ks}=0.01$, respectively; for bipolar and elliptical PNe, $p_{\rm ks}$(Se)~=~0.42 and $p_{\rm ks}$(Kr)~=~0.21.  The low enrichment factors for Type~I and bipolar PNe are likely due to the small masses of their progenitor stars' intershell layers (relative to those of lower mass AGB stars), as well as the severe dilution that the enriched material experiences when it is dredged-up into the massive envelopes of these stars \cite{ll05}.  This result is in agreement with the marginal Zr enrichments found by \cite{gh07} in a sample of intermediate-mass AGB stars.

\begin{figure}[t]
\centering
\includegraphics[height=12cm]{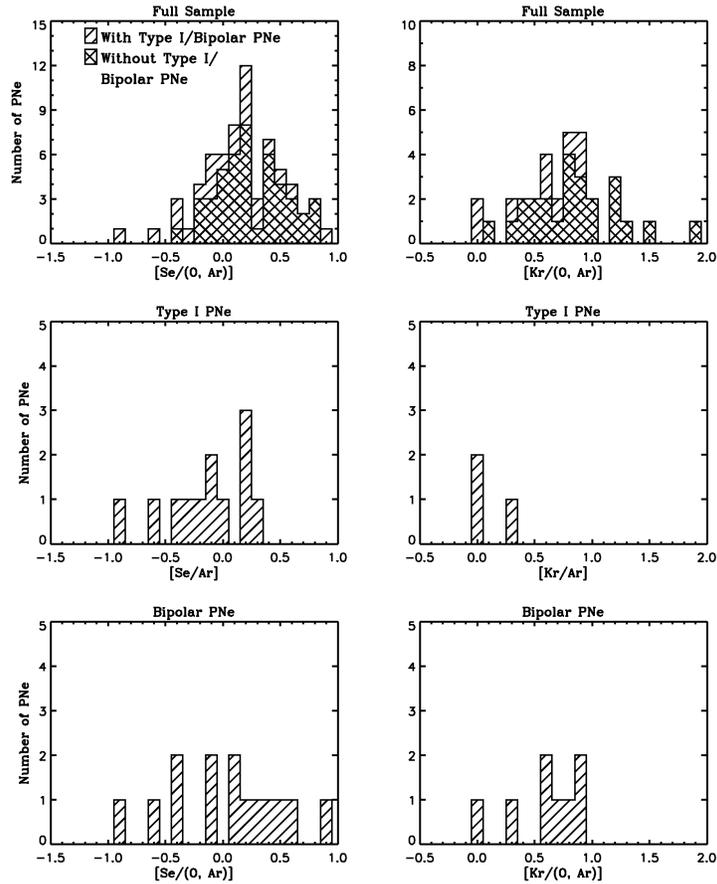}
\caption{Histograms of Se and Kr abundances, separated into 0.1~dex bins, are shown for the full sample (top panels), Type~I PNe (middle panels), and bipolar PNe (bottom panels).}
\label{fig2}
\end{figure}

[WC] PN central stars are H-deficient objects whose surface compositions are enriched in C and (likely) \emph{n}-capture elements \cite{wh06}.  It may therefore be expected that their surrounding nebulae are also enriched in these nuclei.  However, we find that there is no systematic difference between the Se and Kr enrichments of [WC] PNe and nebulae with H-rich central stars (Table~1): KS tests show that there is an 85\% and 99\% probability that the Se and Kr abundances (respectively) in these two PN subclasses are drawn from the same cumulative distribution functions.  It has previously been shown that the compositions of [WC] PNe are similar to PNe with H-rich nuclei for other elements \cite{pena01}, even C \cite{dm01}.

Relatively few PNe in our sample exhibit \emph{direct} evidence for binary central star systems.  Izzard (\cite{izz04}, these proceedings) showed that binary interactions that take place during the thermally-pulsing AGB can truncate this phase of evolution by enhancing the mass-loss of the AGB star.  If this occurs, these systems will be less enriched in TDU products than otherwise similar single stars.  Interestingly, the 14 objects in our sample that display evidence of binary central stars exhibit little to no \emph{s}-process enrichment (Table~1).  However, this is not necessarily due to the binary companions, since the level of enrichment is also strongly dependent on the PN progenitor's initial mass.

\section{Conclusions and Future Work}
\label{sec4}

We have highlighted the major results from our large-scale survey of near-infrared [Kr~III] and [Se~IV] emission lines in Galactic PNe.  We detected Se and/or Kr emission in 81 out of 120 objects in our sample, of which 41 are significantly enriched due to \emph{in situ} \emph{s}-process nucleosynthesis and TDU in their progenitor stars.  Se and Kr abundances are correlated with the gaseous C/O ratio, as predicted by theoretical nucleosynthetic models.  Type~I and bipolar PNe, whose progenitors are IMS, exhibit little if any \emph{s}-process enrichment compared to objects with less massive progenitors.  On the other hand, we find that [WC] PNe are not systematically more \emph{s}-process enriched than objects with H-rich nuclei, which stands in stark contrast to the strong C and probable \emph{s}-process enrichments of [WC] central stars.  PNe with binary central star systems show little evidence of \emph{s}-process enrichments, as may be expected if binary interactions truncated their thermally-pulsing AGB phase.

The derived Se and Kr abundances are uncertain by a factor of 2--3 for most objects in our sample.  We have shown that these uncertainties arise primarily from the ICFs \cite{sdk07}, for two reasons.  First, we have detected only one ion each of Se and Kr.  Therefore, the ICFs can be large and the uncertainties significant.  Secondly, the atomic data controlling the ionization balance of Se and Kr --- photoionization (PI) cross-sections and rate coefficients for various recombination processes --- are poorly if at all known (indeed, this is true of most \emph{n}-capture element ions).  To derive the Se and Kr ICFs, we used approximations for these atomic data in Cloudy and XSTAR \cite{sdk07}.

We have recently instigated a new project to improve the accuracy of \emph{n}-capture element abundance determinations in PNe.  We are observing \emph{s}-process enriched PNe drawn from our near-infrared sample in the optical, in order to search for lines from additional ions of Kr and Se (as well as transitions of other \emph{n}-capture elements).  In the optical spectra of five PNe, we have detected [Kr~IV] in four and [Kr~V] in one thus far.  The detection of additional ions reduces the magnitude and importance of uncertainties in the ICFs.  In addition, we are computing new atomic data for the first six ions of Se and Kr with the atomic structure code AUTOSTRUCTURE \cite{bad86}, including PI cross-sections and rate coefficients for radiative and dielectronic recombination.  These calculations are complemented by experimental absolute PI cross-section measurements near the ionization threshold of these ions, performed at the Advanced Light Source synchrotron radiation facility in Berkeley, CA.  The new atomic data determinations will enable us to derive more accurate and robust ICFs with photoionization models.  Combined with the new observational data, this will significantly reduce uncertainties in Se and Kr abundance determinations in PNe and allow for a more rigorous investigation of \emph{s}-process enrichments.

This work has been supported by NSF grants AST~97-31156 and AST~04-06809.



\end{document}